# Dynamics of laser-induced electroconvection pulses

N. C. Giebink, E. R. Johnson, S. R. Saucedo, E. W. Miles, K. K. Vardanyan, and D. R. Spiegel[*]

*Department of Physics and Astronomy, Trinity University, San Antonio, TX*

C. C. Allen

*Department of Physics, Angelo State University, San Angelo, TX*

## *Abstract*

We first report that, for planar nematic MBBA, the electroconvection threshold voltage has a nonmonotonic temperature dependence, with a well-defined minimum, and a slope of about -0.12 V/degree near room temperature. Motivated by this observation, we have designed an experiment in which a weak continuous-wave absorbed laser beam with a diameter comparable to the pattern wavelength generates a locally supercritical region, or pulse, in dye-doped MBBA. Working 10-20% below the laser-free threshold voltage, we observe a steady-state pulse shaped as an ellipse with the semimajor axis oriented parallel to the nematic director, with a typical size of several wavelengths. The pulse is robust, persisting even when spatially extended rolls develop in the surrounding region, and displays rolls that counterpropagate along the director at frequencies of tenths of Hz, with the rolls on the left (right) side of the ellipse moving to the right (left). Systematic measurements of the sample-voltage dependence of the pulse amplitude, spatial extent, and frequency show a saturation or decrease when the control parameter (evaluated at the center of the pulse) approaches ~ 0.3. We propose that the model for these pulses should be based on the theory of control-parameter ramps, supplemented with new terms to account for the advection of heat away from the pulse when the surrounding state becomes linearly unstable. The advection creates a negative feedback between the pulse size and the efficiency of heat transport, which we argue is responsible for the attenuation of the pulse at larger control-parameter values.

---

[*] To whom correspondence should be addressed.

## I. Introduction

Electroconvection (EC) within nematic liquid crystals has been recognized for some time as a very useful tool for studies of nonequilibrium pattern formation in a dissipative system that is intrinsically anisotropic.[1-3] A wide range of EC investigations have been carried out using planar[4-10] and homeotropic[11-14] nematic alignment; in this work we are interested in the planar configuration. The EC process is in some respects the anisotropic counterpart of Rayleigh-Bénard convection (RBC), with the EC electric field replacing the vertical thermal gradient in RBC.[1] However, the large aspect ratio in EC, along with the small dissipative transport coefficients and the small sample thickness (usually tens of microns), are important experimental advantages for EC that have allowed, for example, recent precision measurements of fluctuations below the onset of a steady-state periodic pattern,[4, 15, 16] along with detailed characterizations of above-onset defect dynamics,[17-19] fluctuations,[7] broken-symmetry modes,[20, 21] and spatiotemporal chaos.[22]

From experimental, numerical, and theoretical perspectives, pattern-forming systems (including EC) have usually been studied in the limit in which the overall size of the complete pattern is much larger than the pattern wavelength.[1] This emphasis is understandable, since the equations of motion are easier to work with when the patterns have a large (ideally infinite) spatial extent. However, there has been a considerable amount of recent interest in *spatially localized* states, known as pulses, in a number of different pattern-forming systems. In general, localized states can arise from either inhomogeneities in the material properties, or from spatially localized solutions to the relevant Ginzburg-Landau (GL) equations within a homogeneous medium. In binary-fluid Rayleigh-Bénard convection, for example, pulse states can be modeled as localized solutions to a quintic GL equation.[23-28] Localized states are also of interest in experimental Turing patterns,[29, 30] Faraday waves,[31] and granular systems,[32] and have been the focus of a number of theoretical and numerical investigations.[33-36] In electroconvection, Dennin and coworkers[22, 37, 38] observed novel localized traveling-wave "worm" pulse states near the EC onset in the nematic liquid crystal I52, with a fixed width in one direction and a much longer and varying length in the other direction. The worms states have attracted considerable attention as an example of spatiotemporal chaotic behavior within localized states near the threshold of linear stablity.[37]

It is thus clear that studies of spatially localized states can lend important new insight into pattern-formation dynamics. With this in mind, we wish to report a new experimental method for the production of a localized supercritical EC region, allowing the creation of an EC pulse with an *experimentally controlled* amplitude and size. A low-power continuous-wave absorbed laser, focused to a spot size that is comparable to the pattern wavelength, produces an elliptical electroconvection pulse with counterpropagating rolls. The sample is nematic dye-doped MBBA held below the laser-free threshold voltage. Unexpectedly, the pulse shape and roll motion are maintained even when the voltage across the sample is large enough to produce spatially extended rolls that surround the pulse. We carry out systematic studies, as a function of the control parameter, of the amplitude, spatial extent, and oscillation frequency of the EC pulses. We find that these properties increase for increasing small values of the control parameter, but saturate or in fact decrease at larger control-parameter values as the amplitude of the spatially extended rolls that surround the pulse increases. We propose that the pulse dynamics can be understood within the framework of ramped pattern-formation theory, supplemented with an important advection-based coupling between pulse size and the thermal transport properties of the liquid crystal. Although we focus here on electroconvection, the fundamental feature of our experimental technique – creation of a locally supercritical region using laser-induced thermal excitation on a length scale comparable to the pattern wavelength – should be readily applicable to other pattern-forming systems. At the same time, we emphasize that the pulses investigated here are of fundamental interest in their own right as dynamic spatially localized states within a pattern-forming nonequilibrium system possessing an intrinsic anisotropy.

The electroconvection work reported here is qualitatively different from the experiments of Dennin and coworkers on worm pulses near threshold.[22, 37-39] The worm states, which to date have only been observed in the liquid crystal I52, arise spontaneously and have very different shapes and dynamic properties compared to the experimentally controlled pulses in MBBA discussed here. In addition, our works differs in a fundamental way both from the EC experiments reported by Joets and Ribotta[40] and the results of Brand *et al.*[41] Joets and Ribotta reported observations of spontaneous small traveling-wave regions near the EC threshold in MBBA, with all the rolls within an inclusion traveling in the same direction, and in which the direction of travel for different inclusions varied randomly to the right or left along the director. Brand and coworkers observed new spatially localized states working at a temperature close to a smectic-nematic phase-transition in the liquid crystal 10E6, in which case the anisotropy in the conductivity was negative, so that the standard model for electroconvection could not be applied.

It should be noted that the use of absorbed light to generate changes in pattern-forming systems has a long history, beginning with the classic white-light RBC experiments by Chen, Whitehead, and Busse.[42, 43] Recent experiments employing this approach include perturbations of Turing structures with temporally periodic illumination,[44, 45] and a novel investigation by Semwogerere and Schatz[46] on the use of spatially extended laser-light patterns to prescribe specific initial conditions onto a Bérnard-Marangoni convection experiment. In EC, Tóth and coworkers[18] recently employed, as a part of their defect-dynamics studies, absorbed laser light to generate defects on an existing spatially extended EC state. In the present report, by contrast, we utilize the absorbed light to create a distinct pulse with moving rolls that is surrounded, at low control-parameter values, by a linearly stable nonconvecting state. Our work is thus very different both in content and objective from the Tóth *et al*. experiments.

## II. Experimental Procedures

The EC cells are constructed following standard procedures[47] using glass slides (Delta Technologies) coated on one side with indium tin oxide. The final usable sample area is about 1 cm$^2$. The sample thickness was measured at $d = 40 \pm 5$ μm. Using interferometry, we found that the deviation from parallelism between the two slides measured along the director was less than 2 x 10$^{-4}$ radians for all cells employed, corresponding to a thickness change along the director of 0.3% or less over the 0.5-mm length of the sample that is viewed with a 10X microscope-objective lens. Samples constructed in this way were generally usable for 1-2 weeks, with changes on the order of 1% per day in the threshold voltage at 70 Hz for stationary Williams rolls with samples held at a fixed temperature over this period. Each of the measurements reported below was obtained in a single day over a duration of 6 hours or less. The nematic liquid crystal was MBBA purchased from Aldrich Inc. The manufacturer lists the purity at 98%; the main impurity is thought to be water, which could allow the Schiff base in MBBA to undergo hydrolysis, resulting in hydrated protons and anionic intermediates. Using the MBBA as received at 70 Hz, we observe, at threshold, normal stationary Williams rolls,[48] which is consistent with the standard model of electroconvection,[2, 49, 50] and with previous experimental studies for samples of approximately the same thickness.[15, 19, 51, 52] The threshold RMS voltage for Williams domains on a laser-free sample at 70 Hz was in the range of 6-7 V for different samples, which is also consistent with MBBA measurements in the literature.[7, 15, 52] To allow absorption at 488 nm, the MBBA was doped with methyl red (MR) at 0.2% by weight, so that about 50% of the laser light was absorbed within the sample. MR is a preferred dopant dye in MBBA transient grating experiments because the two molecules have a similar shape and size.[53] The measured cut-off frequency in undoped samples was on the order of 1 kHz. Using the operational-amplifier circuit discussed by Dennin,[47] we measured the component of the conductivity perpendicular to the plates for the dye-doped MBBA at $(8 \pm 1) \times 10^{-8}$ $(\Omega m)^{-1}$, which was about 10% less than our measured value for the undoped MBBA.

A schematic of the optical apparatus used in our studies is shown in Figure 1. An inverted microscope (Nikon) with a 10X objective lens is equipped with a CCD camera, with a CCD array of 640 x 480 pixels and 8-bit digitization of the intensity at each pixel, for recording the EC shadowgraphs. With a 10X objective lens, the spatial resolution is 1.35 pixels/μm. An image sequence is acquired at rates of 1.7 – 3.8 frames/s and transferred to a computer. The sample is in thermal contact with a copper plate though which water flows to maintain a constant temperature of 24.3 $^o$C. To stabilize the temperature, the sample is housed within a small aluminum box equipped with a glass window for illumination; the temperature stability, measured with a thermocouple at the sample position, was 0.02 $^o$C. The sinusoidal voltage across the sample is provided by a 70-Hz AC power supply that was changed in steps of 50 mV, corresponding to a voltage resolution of about 1%. The sample is illuminated from above with an annular red-LED ring lamp. The 488-nm CW beam from an air-cooled Ar$^+$ laser is directed through the center of the LED ring lamp and focused onto the sample. A long-pass optical filter is inserted between the objective lens and the camera to prevent the transmitted laser light from saturating the CCD array. Polarizer sheets with transmission axes parallel to the nematic director are placed above and below the sample, with a hole cut in the upper polarizer sheet to pass the laser beam. The laser light itself is polarized parallel to the director. Noting that real and virtual images (and combinations of the two) of the convection rolls are formed at various positions along an axis perpendicular to the nematic plane,[54] we consistently positioned the objective lens to view a real image dominated by the fundamental roll wavelength (see Figure 3 below), with minimal contributions from higher-order spatial harmonics. Coordinates are chosen such that the $x$ and $y$ axes are parallel and perpendicular to the nematic director, respectively, in the nonconvecting state. For measurements of the amplitude of the convection rolls, we employ the background-divided dimensionless intensity, defined as[55]

$$I(x, y, t) = \frac{g(x, y, t)}{g_b(x, y)} - 1 \qquad (1)$$

where $g(x, y, t)$ and $g_b(x, y)$ are the sample and background gray-scale values, respectively, obtained at the pixel $(x_i, y_j)$ of the CCD array at time $t$, and where $0 < g < 255$, $1 \leq i \leq 640$, and $1 \leq j \leq 480$. The sample and background images are taken under identical conditions, with the laser present in both cases, except that the voltage across the sample is set to zero for the background image. We have not to applied post-acquisition spatial-frequency filters to any of the images.

### III. Results and Discussion

To explain and motivate our approach to the controlled production of EC pulse states, we show in Figure 2 our result on the temperature dependence of the EC threshold voltage $V_{th}$ in undoped planar nematic MBBA. The sample is 40 μm thick and is subjected to a sinusoidal voltage at 70 Hz. The threshold voltage *vs.* temperature measurement displayed a drift of about -1% per hour. Figure 2 has been corrected for this drift, assuming the drift is linear with time, resulting in a maximum correction of about 5% in $V_{th}$. We note that the portion of Figure 2 showing a decrease in threshold voltage with temperature is consistent with the results of Rehberg et al.[56] It is quite interesting that the temperature dependence is nonmonotonic, displaying a minimum at $T_{\min} = 31.2 \pm 0.5\ ^\text{o}\text{C}$, which is about 7 $^\text{o}$C degrees below the nematic-isotropic transition temperature $T_{NI}$ for this sample. One immediate practical conclusion from Figure 2 is that the thermal stability during an EC experiment can be optimized by working at $T_{\min}$. The possibility of an increase in the threshold voltage as the temperature approaches the nematic-isotropic phase transition can be justified qualitatively on the basis of the standard model for EC,[2, 49, 50] in which the threshold voltage can be expressed in terms of four separate effects.[2] First, a homogeneous director field is destabilized by the product $\tau_q \sigma_a^{eff}$, where $\tau_q$ is the charge relaxation time and $\sigma_a^{eff} > 0$ is an effective anisotropy in the electrical conductivity. Stabilization of the homogeneous state, on the other hand, arises due to (1) an effective elastic constant $K^{eff}$, (2) an effective dielectric anisotropy (in SI units) of $\varepsilon_0 \varepsilon_a^{eff} < 0$, and (3) a viscosity ratio $\eta^{eff}/|\alpha_2|$, where $\eta^{eff}$ is an effective shear viscosity and $\alpha_2$ is the second Leslie coefficient. The standard model shows[2] that if the decrease with temperature of the product $\tau_q \sigma_a^{eff} |\alpha_2|/\eta^{eff}$ is more rapid than the decrease in $|\varepsilon_0 \varepsilon_a^{eff}|$, then it is possible that the nonconvecting homogeneous state will be stabilized (or equivalently, the EC threshold voltage will increase) as the temperature increases.

Since the concentration of conducting impurities in MBBA samples varies widely in the literature,[57] a careful and quantitative evaluation of the minimum in Figure 2 vis-à-vis the standard EC model demands precise measurements of the conductivity both parallel and perpendicular to the director close to $T_{NI}$ using the same samples employed in the EC experiments. We are currently constructing the apparatus required for this type of measurement. For the purposes of the present study, however, the important feature of Figure 2 is that if one works near room temperature, then the dependence of the threshold voltage on temperature is quite significant,[56] with threshold-voltage changes that are about -0.12 V/$^\text{o}$C. This means that with an appropriate temperature spike confined to a small spatial region in MBBA held at a bulk temperature $T$ near room temperature, it should be possible to produce a localized supercritical pulse on a linearly stable uniform background, which is the subject of the current paper. Conversely, however, it should also be noted that if one instead employs a bulk temperature $T > T_{min}$, one should be able to generate a linearly stable nonconvecting "hole" within a spatially extended EC pattern.

Turing now to results in dye-doped MBBA, if a sample held at 24.3 $^o$C is illuminated with the focused laser beam, we indeed find that when the RMS voltage across the sample exceeds a critical voltage $V_c$, an anisotropic (roughly elliptical) supercritical pulse is formed, with the long axis along the director. An example of such a pulse is displayed in Figure 3a, which depicts a background-divided image. The RMS voltage across the sample in Figure 3a is about 17% less than the voltage required for traditional normal-roll formation on the laser-free sample. The cross-section of the laser beam itself, incident at a power of 30 µW, is not readily apparent in this image because the microscope is focused on a real image of the rolls rather than the sample itself, and because most of the laser light transmitted through the sample is extinguished by the long-pass optical filter. The 488-nm laser beam cross-section is approximately Gaussian with a width (FWHM) of about 50 microns, which is comparable to a typical wavelength within the roll pattern. The rolls within the pulse are not stationary, as we show with the spacetime contour plot of Figure 3b, which displays the dimensionless background-divided intensity at a sequence of different times. To generate Figure 3b, an intensity trace $I(x, t)$ is obtained for each image by averaging the background-divided intensity $I(x, y, t)$ over the $y$ coordinate over 16 pixel-rows (a width of $\Delta y = 12$ µm, as shown in Figure 3a) centered vertically on the EC pulse. As seen in Figure 3b, we observe counterpropagating rolls on opposite sides of the ellipse, with velocities on the order of 0.1 – 0.2 wavelengths/sec. As the counterpropagating roll patterns collide at the center of the ellipse, a roll is periodically annihilated and then replenished at the center. We should note that when the sample is initially illuminated with the laser, we use neutral density filters to adjust the laser power so that the pulse size appeared to neither increase or decrease with time; our best results were obtained following a continuous illumination for about one hour with a laser power in the range of 30 µW – 120 µW. With a proper choice of laser power, we found that an EC pulse of constant size could be maintained at constant voltage and constant laser power for at least 10 hours, which is equal to several thousand of the oscillation periods (to be discussed quantitatively below) evident in Figure 3b. It is important to note that, although the pulse is centered on the laser-illuminated region, the spatial extent of the pulse in Figure 3a is significantly larger than the laser-beam cross-section.

If the voltage across the sample is increased, we observe that the size of the counterpropagating-roll pulse region initially increases. With continued increases in the voltage, we observe a spatially extended Williams-roll region *outside* the central pulse region, moving with a significantly slower speed. As these slower "outer" rolls become more pronounced with further increases in voltage, the central pulse region shows a *decrease* in size. The background-divided image and the spacetime profile for an image sequence in which the outer rolls are clearly visible are shown Figures 3c and 3d, respectively. The image sequence used for Figures 3(c-d) was obtained with a sample held at a voltage that was about 9% less than the voltage required for normal roll formation on the laser-free sample. Figure 3d shows that while both the "central" rolls and the "outer" rolls on the left and right move toward the center, the speed of the outer rolls is slower by about an order of magnitude. The presence of the outer, slower-moving spatially extended state can be understood as the result of a temperature increase (and, in accordance with Figure 2, a corresponding decrease in threshold voltage) due to the diffusion of heat from the laser-excitation region into the surrounding region. It is important to note, however, that a well-defined localized pulse state persists, with a characteristic roll speed on the order of 0.1 wavelength/s, even after the homogeneous state outside the pulse region becomes linearly unstable. We will henceforth refer to the central spatially localized region of counterpropagating rolls as a "counterpropagating anisotropic pulse state" (CAPS), and to the slower spatially extended outer convection pattern (if present) as the "outer Williams state".

The laser polarization direction was parallel to the director for the images shown in Figure 3. However, we found that the CAPS was also formed when the laser polarization was rotated with a half-wave plate to be orthogonal to the director. This implies that it is very likely that it is the heat dose from the laser beam, rather than the well-known nonlinear optical properties of dye-doped nematics,[58] that is responsible for the formation of the CAPS. Further evidence for a thermal origin of these states was gained from an experiment in which we replaced the methyl red dye with 1% anthraquinone by weight. The laser power was increased to 3 mW because anthraquinone has a much smaller absorption coefficient at 488 nm than methyl red. We found the counterpropagating pulse state was again produced under these conditions, demonstrating that the detailed molecular structure of the dye does not appear to be a critical factor. The data shown in Figure 2 can be used to calculate an upper bound on the temperature difference between the center of the pulse and the bulk held at 24.3 °C. Since EC first occurs, as the voltage is increased, at the *center* of the illumination region where the temperature $T_0$ is a maximum, we must have $T_0 < T_{min}$, where the latter temperature refers to the minimum in Figure 2. The temperature difference generated by the laser heating is therefore not larger than $T_{min} - 24.3 \,°C \approx 7 \,°C$.

To enable a quantitative description of the CAPS dynamics, we first discuss systematic measurements of the mean-square amplitude of the moving rolls within the CAPS region as a function of the mean-square voltage across the sample. A background-divided intensity trace $I(x, t)$ along the director is obtained from a 200 x 16-pixel central CAPS area (about 150 μm x 12 μm along *x* and *y*, respectively) by averaging over the *y* direction in the manner discussed above in the context of Figure 3. The mean-square dimensionless intensity $\langle I^2 \rangle$ is then obtained from averages over time and over the *x*-coordinate within this area. At each voltage, a sequence of 260 images was obtained at a rate of 1.7 images/s. The voltage was changed in 50-mV steps, and there was a 10-minute rest interval after the voltage was changed before a new image sequence was acquired. A noise level of $\langle I^2 \rangle_{noise} = 2.7 \times 10^{-4}$ was obtained from a background-divided measurement of $\langle I^2 \rangle$ within a 200 x 16-pixel area well outside the CAPS region, and was subtracted from all mean-square intensities recorded within the CAPS region. A plot of $\langle I^2 \rangle$ vs. the mean-square voltage across the sample is shown in Figure 4a. We observe that $\langle I^2 \rangle$ increases at low voltages, achieving a maximum near $V^2 = 39 \,V^2$, and then, intriguingly, saturates and falls off slightly at higher voltages. To relate the mean-square amplitude to the square of the director distortion angle $\theta^2$, we employ the relation derived by Rehberg and coworkers:[59]

$$\theta^2 = 0.372 \, (\lambda/d)^2 \, I^2 \qquad (2)$$

where the numerical coefficient applies to MBBA.  In the present study, the conversion from $\langle I^2 \rangle$ to $\langle \theta^2 \rangle$ is difficult for several reasons.  First, as is clear from Figure 3, the CAPS is characterized by a range of wavelengths, rather than a single wavelength, and this wavelength range changes with time as a roll is repeatedly annihilated and then replenished at the CAPS center.  Second, as noted above, the uncertainty in the sample thickness *d* is on the order of 13%.  Finally, in their recent physical-optics study of shadowgraphs, Trainoff and Cannell[60] pointed out that Equation (2) is based on application of Fermat's principle to the extraordinary index function, rather than the extraordinary-ray index function.  Keeping these various limitations in mind, we estimate that the mean-square distortion angle corresponding to the maximum mean-square intensity in Figure 4a is $0.02 \pm 0.01$ rad$^2$, which corresponds to a maximum RMS director distortion of $140 \pm 40$ milliradians.  This value is comparable to amplitudes reported for spatially extended MBBA rolls.[15]

The definition of a control parameter $\varepsilon$ for the CAPS state requires some care.  The control parameter is a clearly a function of position:  for example, when the CAPS region is embedded within a linearly stable homogenous background (as in Figures 3a and 3b), the control parameter varies from $\varepsilon = 0$ at the edges of the CAPS region to some maximum $\varepsilon_{max}$ at the center of the region.  Figure 4b shows the data of Figure 4a at the lowest sample voltages; seven of these points can be fit to a straight line that extrapolates to zero amplitude at a mean-square voltage of $V_c^2 = (24.1 \pm 0.1)$ V$^2$.  The linear dependence is maintained over these 7 points as the mean-square amplitude changes by a factor of about 5.  The critical mean-square voltage $V_c^2$ shown in Figure 4b is evidently the voltage at which the CAPS amplitude goes to zero at the pulse center (where the control parameter is a maximum), so that we may define $\varepsilon_{max} = (V/V_c)^2 - 1$.  This control parameter is shown on the upper axis of Figure 4a.  The linear relation between the control parameter and the mean-square amplitude (or the square root of the control parameter and the RMS amplitude) shows that, within our instrumental resolution, the bifurcation to spatially localized rolls is supercritical, which is consistent with the standard model for EC in MBBA.[2, 49, 50]

The two points at voltages for which $V < V_c$ (or equivalently $\varepsilon_{max} < 0$) in Figure 4b strongly depart from the other points at low voltages.  The mean-square amplitude for these two points is only about a factor of 3 above the noise. The image sequences for these two points, while noisy, indicate that the localized roll structure is *stationary*, in contrast to the counterpropagation observed for all image sequences with $\varepsilon_{max} > 0$. It is thus clear that these two points do not correspond to CAPS states. It should be noted that the mean-square amplitude is not expected to go to zero at $\varepsilon_{max} = 0$ due to subcritical thermal fluctuations in the director alignment. Rehberg *et al.*[15] have found that spatially extended fluctuations in MBBA have a mean-square value at $\varepsilon = -0.01$ that is about an order of magnitude smaller than the mean-square amplitude for $\varepsilon_{max} < 0$ in Figure 4. Fluctuations in the present study, however, take place in a system that is spatially localized, rather than extended, and in the future it would seem worthwhile to apply the methods developed for studies of subthreshold EC fluctuations[4, 15, 16, 55, 59, 61] to the present system for $\varepsilon_{max} < 0$.

Next, we turn to systematic measurements of the spatial extent of a CAPS region. A simple determination of the spatial extent from a measurement of amplitude *vs.* position will not suffice, since such a procedure cannot distinguish between the slow Williams rolls outside the CAPS region and the counterpropagating CAPS rolls when the latter are embedded within the former, as shown in Figure 3(c-d). We have determined a CAPS spatial extent using the time-averaged time derivative of the dimensionless intensity rather than the intensity itself, since the time derivative will be significantly larger within the CAPS region. We thus define a spatial extent $\xi$ as the sum (over $x_i$) of the time-averaged derivative, normalized to its maximum value:

$$\xi = \frac{a \sum_{x_i} \left\langle \left| \frac{dI}{dt} \right| \right\rangle_{x_i}}{\left\langle \left| \frac{dI}{dt} \right| \right\rangle_{max}} \qquad (3)$$

where in this case <> indicates a time average, and *a* is the length of a pixel. To improve the signal-to-noise ratio of <|dI/dt|>, at the cost of some temporal resolution, we first smooth the intensity level at each position $x_i$ by averaging over four consecutive points in time. The absolute value of the time derivative is then obtained from the intensity difference between consecutive smoothed points, and is averaged over time. Finally, the sum over $x_i$ in Equation 3 is carried out over all values of $<|dI/dt|>$ that exceed a predetermined noise level $\left\langle \left| \frac{dI}{dt} \right| \right\rangle_{min} > 0.04$. Increasing $\left\langle \left| \frac{dI}{dt} \right| \right\rangle_{min}$ to 0.05 did not have a noticeable effect on the calculated spatial extent. If instead $\left\langle \left| \frac{dI}{dt} \right| \right\rangle_{min}$ was reduced to 0.01, the calculated spatial extent displayed a poor signal-to-noise ratio due to fluctuations in the calculated derivative. In Figure 5 we show the dependence of the spatial extent on the control parameter $\varepsilon_{max}$. The images used for these measurements are the same as those used in the amplitude measurement discussed above. The three points shown at the lowest $\varepsilon_{max}$ had values for $<|dI/dt|>$ that were below the noise level at all $x_i$. The value (with uncertainty) of $\varepsilon_{max}$ at which the outer Williams rolls first appear is indicated as a bar at the bottom of the plot. The spatial extent $\xi$ increases to a maximum near $\varepsilon_{max} = 0.4$, and then falls off significantly at greater $\varepsilon_{max}$. The decrease in $\xi$ at higher $\varepsilon_{max}$, which is also clear from direct inspection of spacetime diagrams such as that shown in Figure 3d, occurs as the outer Williams state grows in amplitude.

    To gain additional insight on the CAPS dynamics, we have also measured the frequency of the oscillations at the center of the CAPS region. For each image frame, we averaged the gray-scale values in a 4 x 4-pixel region (about 3 μm x 3 μm) at the CAPS center, and then determined the oscillation frequency from a fast Fourier transform over time. The results for the dependence of frequency on $\varepsilon_{max}$, using the same image sequence employed for Figures 4 and 5, are shown in Figure 6 up to about $\varepsilon_{max} = 0.45$. We were unable to obtain usable data at higher $\varepsilon_{max}$ because defects moving within the CAPS region at higher voltages clearly distorted the frequency measurements. We see from Figure 6 that the frequency increases with increasing voltages for $\varepsilon_{max} < 0.28$, but the increase is not dramatic: the frequency changes by about 20% as $\varepsilon_{max}$ changes from 0.015 to about 0.28. Like the CAPS amplitude and spatial extent, the frequency displays a maximum value as the control parameter is varied; however, it is important to note that the frequency does not appear to continuously approach

zero as $\varepsilon_{max} \to 0$, in contrast to the amplitude and spatial extent. As shown with the bar at the bottom of the Figure 6, the maximum frequency occurs within the range of $\varepsilon_{max}$ in which the outer Williams rolls first appear. Because the frequency does not change by a large percentage as $\varepsilon_{max}$ is varied, drift of material parameters (if present) during the measurement could result in large errors. To test for drift, we measured the frequency dependence of a different sample at a number of different voltages ($V_{A1}$, $V_{A2}$, $V_{A3}$,…), and then immediately employed intermediate voltages $V_B$ such that $V_{A1} < V_{B1} < V_{A2}$, $V_{A2} < V_{B2} < V_{A3}$, …etc. We found that, to within a frequency uncertainty of a few percent, the $V_A$ and $V_B$ sets fell on the same curve, which demonstrates that the change in frequency shown in Figure 6 is not a result of experimental drift. Although the basic shape of the frequency dependence on $\varepsilon_{max}$ shown in Figure 6 was observed in a number of different samples, we found that the maximum in the absolute frequency varied over an approximate range of 0.1 Hz – 0.2 Hz for different samples prepared in the same manner, as exemplified in a comparison of Figure 3b *vs*. 3d.

Since the laser-induced pulse state displays moving rolls, it is important to investigate how the roll motion is affected by changes in the laser intensity. We found that when neutral density filters were used to reduce the laser power by a factor of 4, the frequency was reduced only by about 10%, and the wavelength was not altered by a measurable amount.

As noted in the Introduction, localized states can arise from either from material inhomogeneities, or from spatially localized solutions to the appropriate GL equations applied to a homogeneous medium. Two additional experiments were carried out, each designed to test the importance of material inhomogeneities.

> (I) We first illuminated the sample with a 1-mW laser beam for 25 s while holding the voltage across the sample at zero. The application of the voltage across the sample, which was in this case 4% lower than the laser-free threshold voltage, was *delayed* until 10 s after the laser beam was shuttered. We found that, even after the delay, the CAPS state still formed in the previously illuminated region. There is a complete absence of any pattern-formation dynamics (due to the lack of a sample voltage) during the laser illumination, but the material inhomogeneity generated by the absorbed laser light evidently allows a CAPS pulse centered on the previously illuminated area to form once the voltage is applied.

(II) Next, we used cylindrical optics to elongate the laser spot into a rectangle measuring about 300 μm x 50 μm, with the long side perpendicular to the rolls. The sample was held at a voltage that was 4% less than the laser-free threshold voltage. The laser power was increased to 3 mW, and we used a rectangular aperture to clip the tails of the laser-beam cross section and improve the uniformity of the laser intensity. A direct image (with no background division), and the resulting spacetime diagram in which we again averaged the intensity over $\Delta y = 12$ μm, are displayed in Figures 7a and 7b, respectively. We again observe counterpropagating rolls moving along the director axis, but in this case the motion is confined to the edges of the rectangular laser region that are perpendicular to the director, where the laser-intensity gradients (ramps) are largest, and these ramped regions are separated by a nearly stationary set of rolls (with a number of dislocation defects apparent in the bright-line region) within the central illumination area.

The experiments (I) and (II) indicate that the material inhomogeneity introduced via laser illumination is a fundamental feature of the CAPS, and that the roll motion is driven by gradients in the laser intensity, which create ramps in the material properties. It is therefore unlikely that the CAPS can be explained with the type of Ginzburg-Landau models with spatially uniform material properties that have been successfully applied to binary fluid convection.

The couterpropagating-roll dynamic is an intriguing feature of the CAPS. In general, one expects drift in the presence of gradients caused by inhomogenities,[51] and Figure 7 shows that the CAPS motion is indeed gradient-driven. There are several findings, however, which indicate that a quantitative model for the CAPS must consist of more than simple drift due to laser-based inhomogeneities. First and most importantly, a localized CAPS state with a well-defined characteristic roll speed is maintained even when the heat dose from the laser creates "outer rolls" that surround the pulse (Figures 3c and 3d). Second, systematic variation of $\varepsilon_{max}$ results in an increase in the CAPS amplitude, spatial extent, and oscillation frequency for small $\varepsilon_{max}$, but all three parameters saturate or decrease when $\varepsilon_{max}$ reaches ~ 0.3 (Figures 4 - 6). Finally, the frequency (hence the speed) of the roll motion does not vanish for small $\varepsilon_{max}$, and changes only slightly even when the laser intensity (hence the absolute laser-intensity gradient) is reduced by a factor of four. Thus although the experiments show that the basic driving force for the CAPS motion is the laser-induced thermal gradients, our results also necessitate a model for the CAPS that goes beyond simple inhomogeneity-induced drift.

With this in mind, a promising approach to modeling the CAPS, in our view, is based on the theory of control-parameter ramps, initiated in 1982 by Kramer and coworkers[62] as a mechanism for wavelength selection in nonequilibrium pattern formation. In this work, the authors used a nonlinear phase-diffusion equation to examine the effects of a control parameter that varied slowly with position (caused, for example, by ramps in the material properties) within a reaction-diffusion system. The fundamental result is that a sub-to-supercritical spatial ramp in the control parameter will result in a unique function $q(x)$ for the wavenumber. These ideas have been extended by Riecke and Paap,[63, 64] who noted that when $q(x)$ persists into the Eckhaus-unstable range, traveling waves will be generated in the ramped regions. The predictions of Riecke and Paap were carefully tested in experiments on ramped Taylor vortex flow (TVF) carried out by Ning, Ahlers, and Cannell,[65] and excellent agreement between theory and experiment was obtained.

The CAPS dynamics, especially as depicted in Figure 7, display significant qualitative similarity to the ramped TVF experiments. In both the TVF and CAPS experiments, moving rolls propagate up a ramp from low to high values of the control parameter, towards a uniform region in which the rolls are stationary. A control-parameter-ramp model also provides an immediate explanation for the maximum in the oscillation frequency, which is observed both for ramped TVF and for CAPS: as the control parameter is increased, a wider bandwidth of stable states is available to accommodate the required $q(x)$, which decreases the coupling to Eckhaus-unstable wavenumbers and hence reduces the oscillation frequency. Thus, it is possible in Figure 7 that the CAPS rolls moving toward the center compress the uniform region, driving it into an Eckhaus-unstable state, which in turn drives more rolls up the ramp. The CAPS produced with Gaussian laser cross-sections (see Figure 3) would then be essentially the same as the elongated region in Figure 7, except that the "homogeneous" section is collapsed to small region at the center of the beam. In the case of the Gaussian laser cross-section, in which the CAPS size can be comparable to the pattern wavelength $\lambda$, the experiments may not fulfill the slowly-varying condition employed in control-parameter ramp theory. Nevertheless, the similarity that we observe in the behavior of pulses of size $\sim \lambda$ and pulses that are up to a factor of 5 larger (see Figure 5) indicates that the approximation of a slowly varying ramp would not be an unreasonable starting point.

While a control-parameter-ramp model may provide an account for the cause of the roll motion and the observed frequency behavior, this mechanism does not provide an obvious explanation for the saturation in the CAPS amplitude or the rather dramatic decrease in spatial extent (Figures 4 and 5, respectively). A key observation for understanding these effects is that the CAPS amplitude and size attenuate when the "outer" rolls become prominent. As the outer rolls at the CAPS boundary increase in amplitude, the ability of the liquid crystal to conduct heat away from the CAPS region is strongly enhanced by the velocity field that results from convection within the outer rolls (advection heat transport). The effectiveness of the advection process is demonstrated experimentally in Figure 8. For this measurement, cylindrical optics were used to shape the laser cross-section (with a power of 290 µW) into a large-aspect-ratio rectangle (14 µm x 1.5 mm FWHM) with its long side *parallel* to the rolls. The sample was held at a temperature that was 2.3 $^{\circ}$C less than the temperature of the nematic-isotropic transition. At the lowest sample voltage shown, the laser intensity is sufficient to drive a portion of the illuminated region into the isotropic phase, evidenced by the isotropic "bubbles" in Figure 8. As the sample voltage is raised and convecting rolls begin to envelope the illuminated region, the amount of isotropic-phase material is depleted until only the nematic phase is present at the highest sample voltage employed. Returning the sample voltage to a lower value resulted on the reappearance of isotropic bubbles. A simple interpretation of Figure 8 is that the enhancement in thermal transport resulting from the onset of convecting rolls adjacent to the illuminated area reduces the temperature spike produced by the laser. Similarly, an explanation for the CAPS amplitude saturation and size decrease is that the temperature spike produced by the laser beam is reduced via advection of heat when convection rolls are present outside the pulse region, which in turn reduces $\varepsilon_{max}$ and the size of the supercritical region. This is consistent with our observation that the CAPS amplitude and size increase as the control parameter increases until the rolls generated outside the CAPS acquire a significant amplitude, which then precludes a further increase of the pulse amplitude and size. The advection is also expected to reduce the slope of the material-property ramps created by the temperature spike.

A modeling approach for localized states based on ramped standard-model EC equations of motion, supplemented with terms to describe the control-parameter-dependent advection of heat in the presence of a horizontal temperature gradient, could prove very interesting vis-à-vis the experimental results reported here. Heat transport due to advection plays a fundamental role in RBC in isotropic materials,[1] and is necessarily an important aspect of thermally driven convection in nematic liquid crystals.[49, 66-68] Since EC, however, is generally modeled as an isothermal process, heat transport due to advection has not yet been applied (to our knowledge) to EC pattern-formation dynamics. Our experiments motivate a need to incorporate advection heat transport explicitly into the EC equations of motion to develop a quantitative CAPS model. One important open question that could be addressed with such a model is whether the robust persistence of the CAPS, as defined by an abrupt boundary between the pulse and the slow outer rolls (Figure 3c-d), could result from the competing effects of a positive control parameter and advection heat transport. It is also important in future work to test the proposed coupling between pulse size and heat advection with a direct *independent* measurement of the exact temperature profile produced by the absorbed laser both inside and outside of the CAPS region as a function of $\varepsilon_{max}$. We are currently working in our laboratory on both the development of a model incorporating ramps and advection into the EC equations of motion, and the design of experiments for the precise measurement of the temperature profile.

To conclude, after reporting on the temperature dependence of the EC threshold voltage in MBBA, we described the use of absorbed laser light to produce a counterpropagating anisotropic pulse state (CAPS). The CAPS persists even in the presence of spatially extended rolls that surround the pulse. The CAPS amplitude and size increase with increasing control parameter at small $\varepsilon_{max}$; however, the increases are attenuated at larger $\varepsilon_{max}$. We have proposed that the correct quantitative model for the CAPS should be based on a control-parameter ramp model supplemented to include advection heat transport generated by the convection rolls that surround the CAPS at higher $\varepsilon_{max}$. The advection process creates a rather interesting negative feedback between the pulse size and the efficiency of heat transport away from the pulse.

## V. Acknowledgements

We wish to express our gratitude to Guenter Ahlers, Eberhard Bodenschatz, Stephen Morris, and Hermann Riecke for their insightful ideas on these experiments, which were of great value in this work. We are grateful to Ingo Rehberg for sharing data on the temperature dependence of the EC threshold voltage. We also thank Bill Collins, Michael Dennin, Albert Libchaber, Benjamin Plummer, and Harry Swinney for useful discussions and e-mails. We thank Truc Bui-Thanh, Landon Nemoto, and Alan Puchot for their help. Finally, we are grateful to Trinity University and specifically to President John Brazil for encouragement and support. This work was supported in part by a grant from the National Science Foundation (Research at Undergraduate Institutions #0111143). Acknowledgement is also made to the donors of the Petroleum Research Fund, administered by the American Chemical Society, for partial support of this research.

**Figure Captions**

1. Schematic of the apparatus. The vertical gray line indicates the path of the laser beam. A. LED ring lamp. B. Polarizer. C. 50-mm focal length lens. D. Sample housing. E. Sample with leads. F. Sample stage. G. Polarizer. H. Copper plate with water flow. I. Objective lens. J. Long-pass optical filter. K. CCD camera.

2. Temperature dependence of the threshold voltage $V_{th}(T)$ for an undoped MBBA sample.

3. (a) Background-divided image of a laser-induced CAPS pulse embedded within a uniform nonconvecting background, normalized to an 8-bit gray-scale for display purposes. The two short white-line segments on the left axis indicate the set of rows averaged together to obtain the intensity trace $I(x, t)$. The white bar in the lower-left corner represents 100 μm. (b) Spacetime diagram of the pulse in Figure 3a showing counterpropagation toward the center. Time advances upward, and the total time elapsed is 33 s. (c) Background-divided image from a different sample of a pulse surrounded by slow Williams rolls, normalized to an 8-bit gray-scale for display purposes. The two short white lines on the left axis indicate the set of rows averaged together to obtain the intensity trace $I(x, t)$. (d) Spacetime diagram showing the CAPS motion and the slower motion of the outer Williams rolls of Figure 3c. Time advances upward, and the total time elapsed is 66 s. Note the abrupt boundary between the pulse and the outer rolls in Figure 3d.

4. (a) The dimensionless background-divided mean-square intensity plotted as a function of the mean-square voltage across the sample. The upper axis shows the control parameter $\varepsilon_{max}$. (b) Expanded figure showing the points with the smallest mean-square intensity. The best-fit line intercepts the abscissa at $V_c^2 = 24.1 \pm 0.1$ V$^2$.

5. CAPS spatial extent $\xi$ measured as a function of the control parameter $\varepsilon_{max}$. The horizontal bar at the bottom of the graph indicates the range of $\varepsilon_{max}$ over which the outer Williams rolls first appear.

6. Dependence of frequency at the CAPS center on $\varepsilon_{max}$. The horizontal bar at the bottom of the graph indicates the range of $\varepsilon_{max}$ over which the outer Williams rolls first appear.

7. (a) Direct image (with no background subtraction) of the pattern obtained when a cylindrical lens is used to extend the laser cross-section in the $x$-direction. The two short white-line segments on the left axis indicate the set of rows averaged together to obtain the intensity trace $I(x, t)$. The white bar in the lower-left corner represents 100 µm. (b) Spacetime diagram of the pattern shown in Figure 7a. Time advances upward, and the total time elapsed is 155 s.

8. Direct images showing EC suppression of isotropic bubbles. The voltage across the sample in each case is indicated.

**Figure 1**

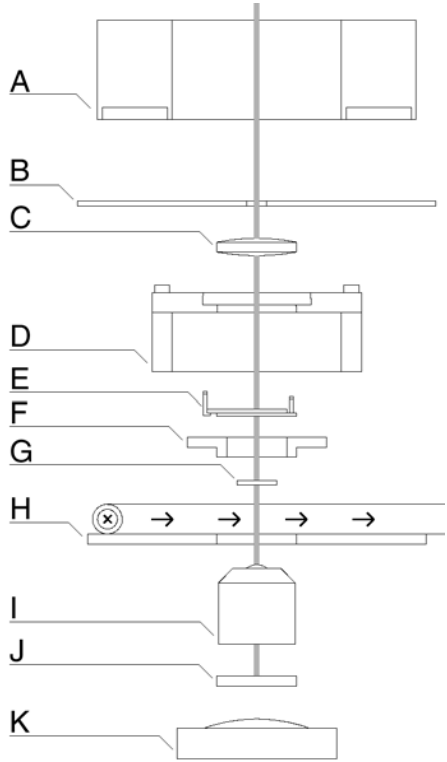

**Figure 2**

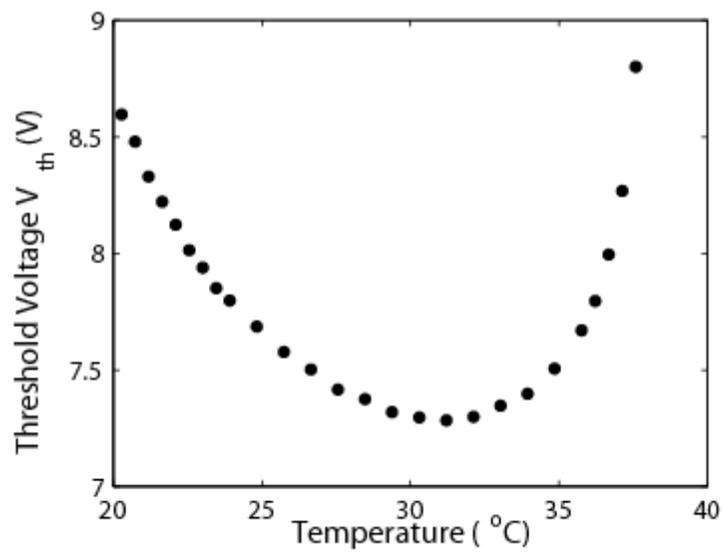

**Figure 3**

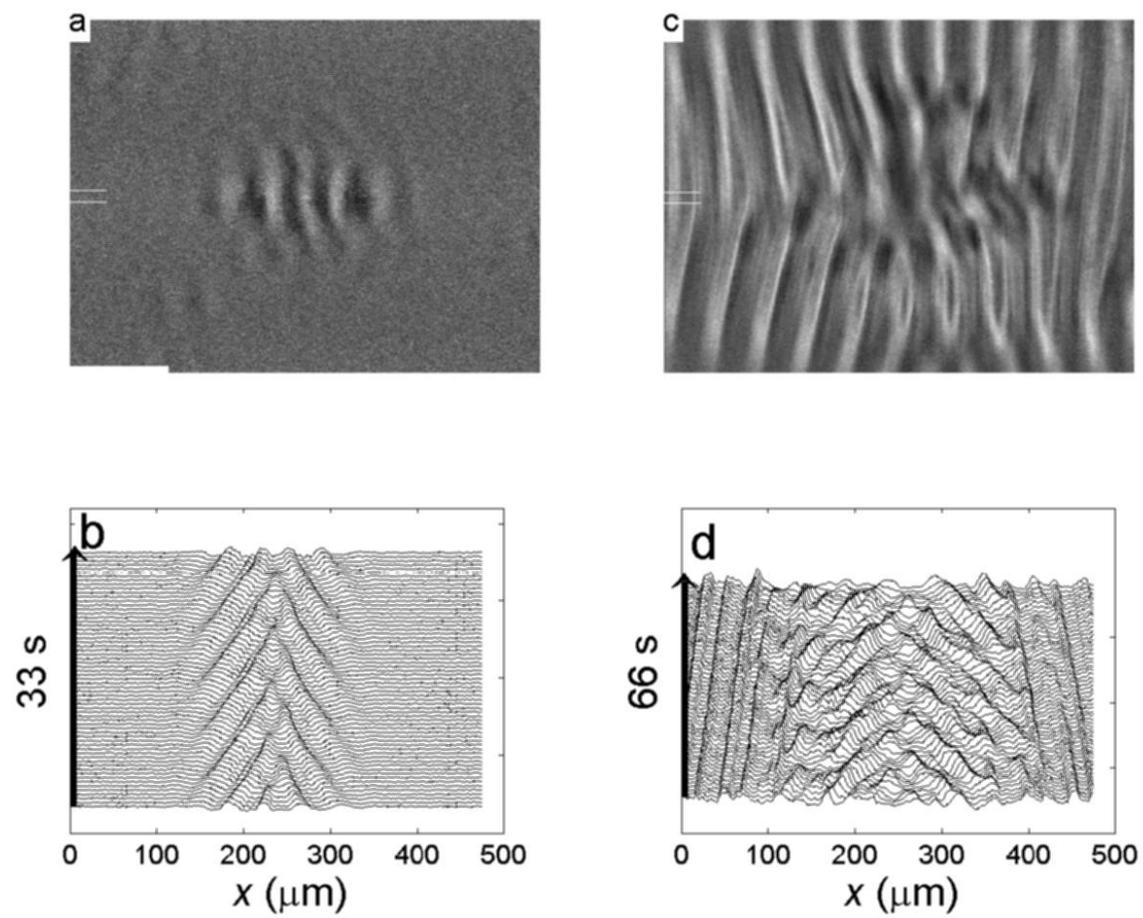

**Figure 4a**

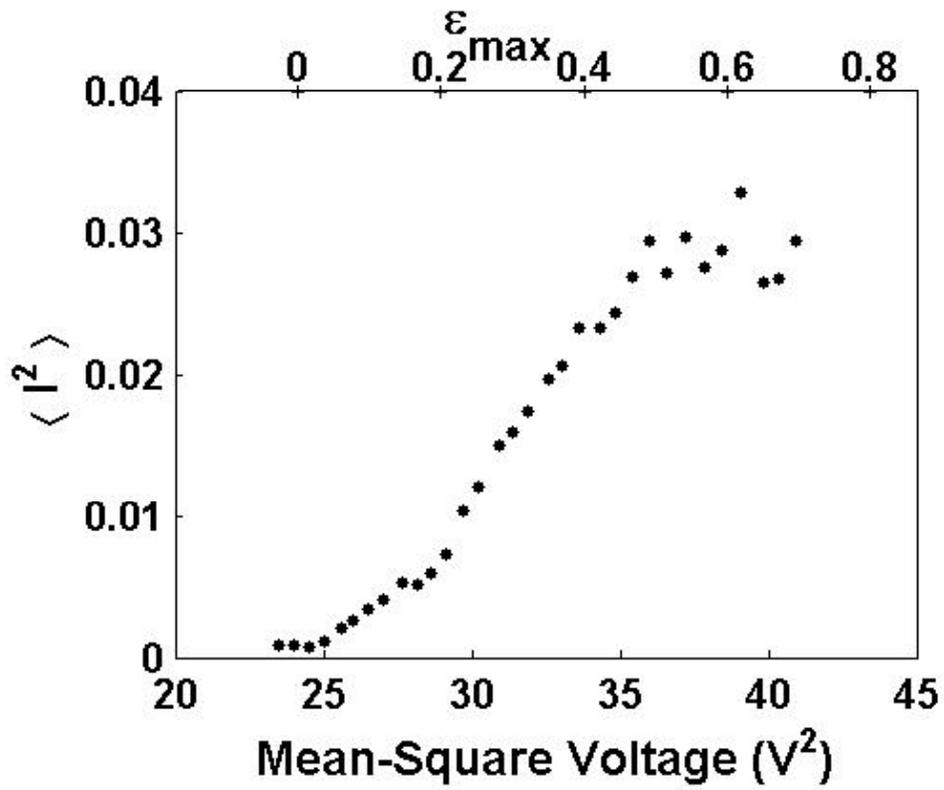

**Figure 4b**

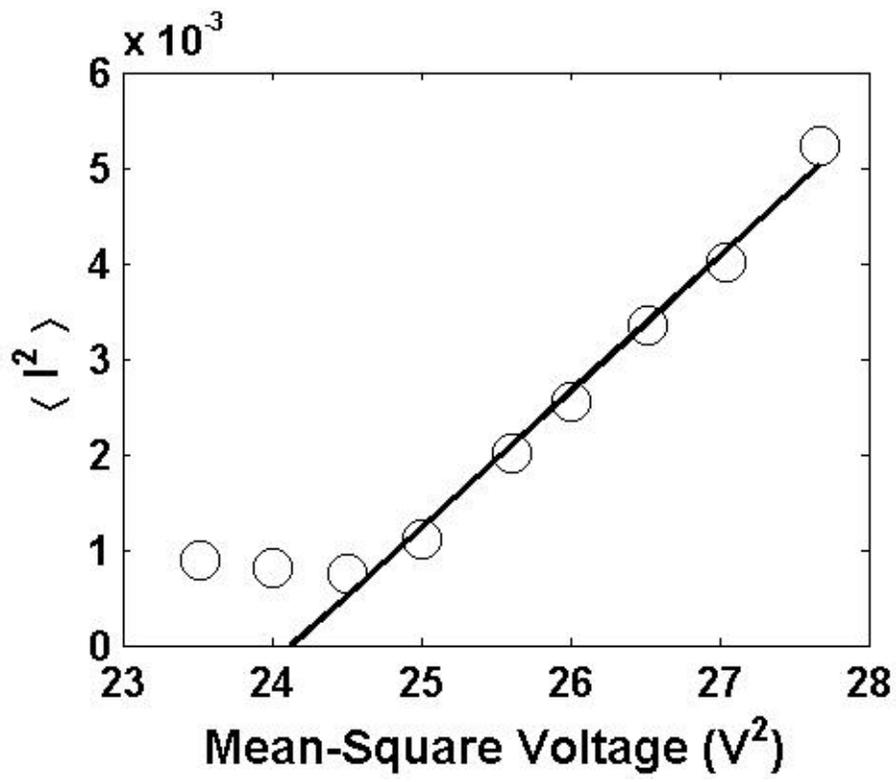

**Figure 5**

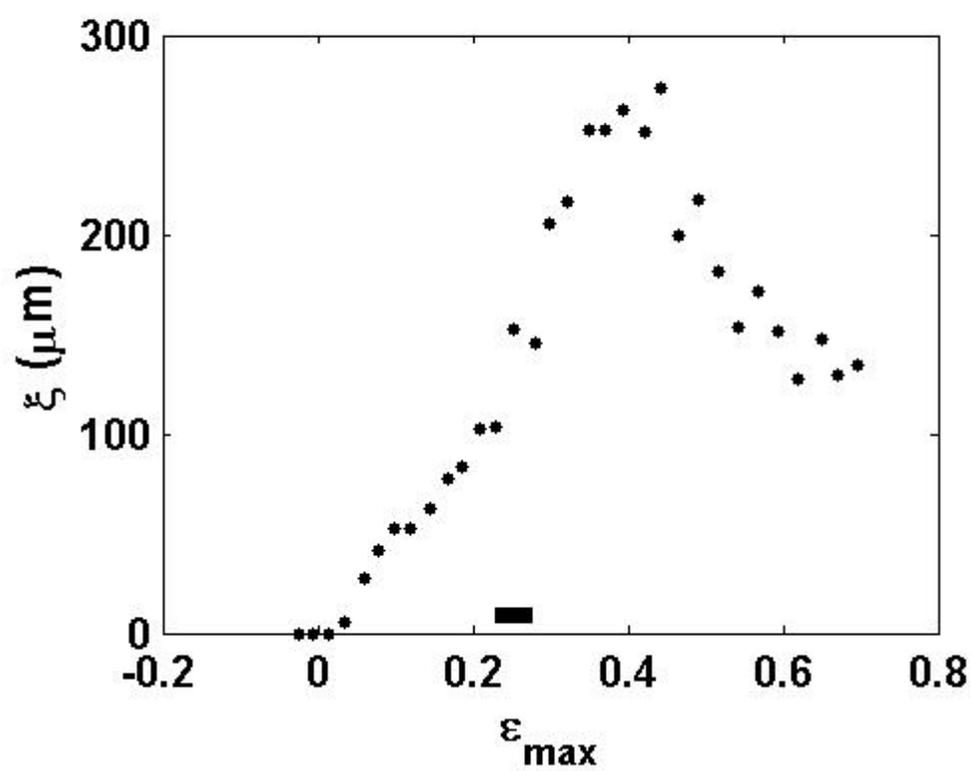

**Figure 6**

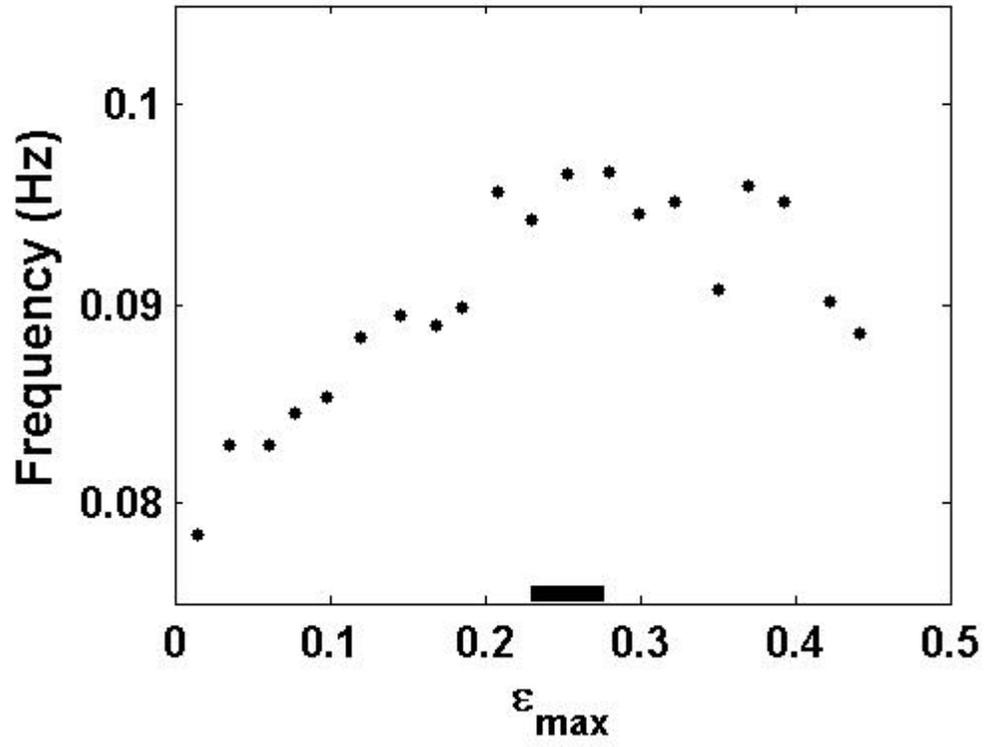

**Figure 7a**

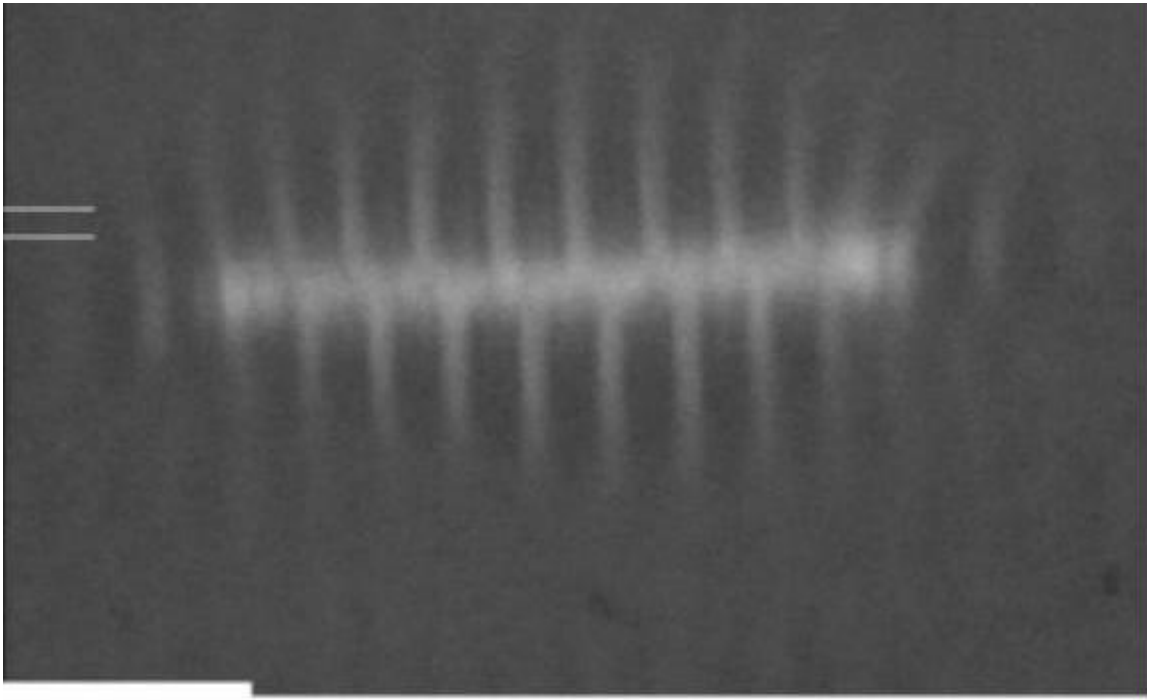

**Figure 7b**

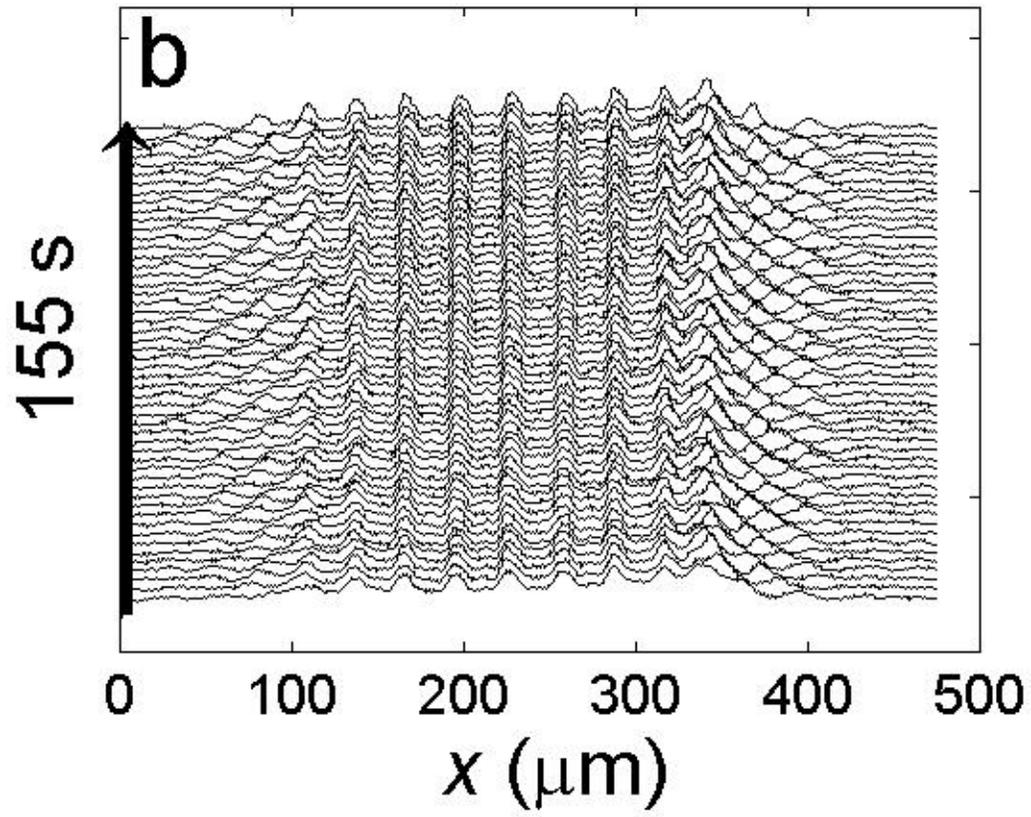

**Figure 8**

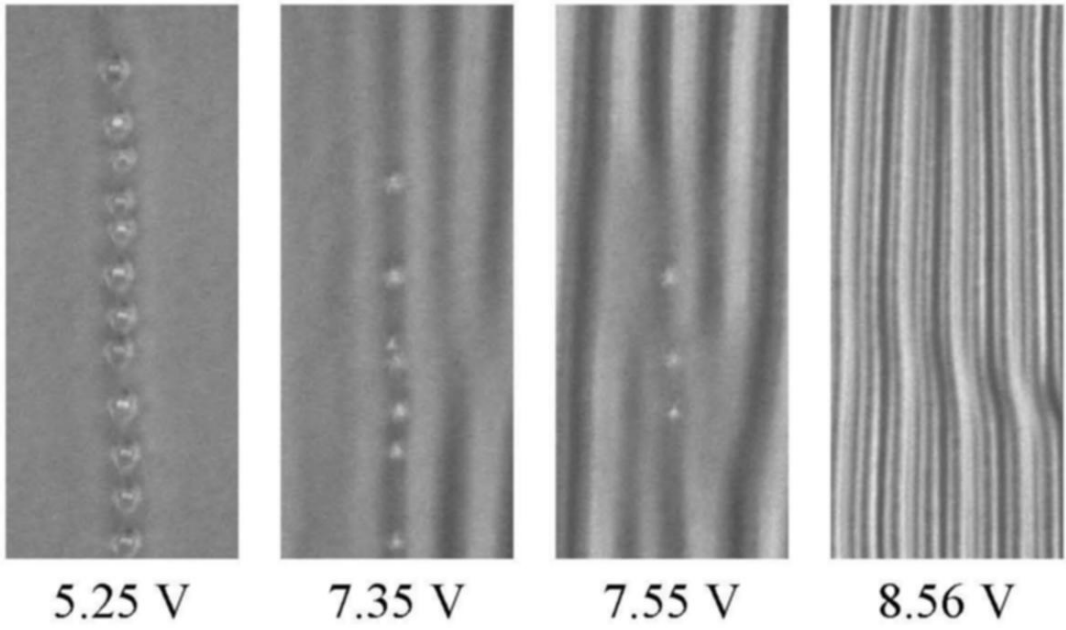